\documentclass[fleqn,usenatbib]{mnras}

\usepackage{amssymb}	
\usepackage{newtxtext,newtxmath}

\usepackage[T1]{fontenc}
\usepackage[utf8]{inputenc}

\DeclareRobustCommand{\VAN}[3]{#2}
\let\VANthebibliography\thebibliography
\def\thebibliography{\DeclareRobustCommand{\VAN}[3]{##3}\VANthebibliography}

\usepackage{graphicx}	
\usepackage{amsmath}	
\usepackage{enumitem}
\usepackage{booktabs,tabularx,graphicx,amsmath,makecell,hhline,enumitem,gensymb,multirow,ulem,adjustbox}
\usepackage{MnSymbol,bbding,pifont}
\usepackage{hyperref} 
\usepackage[dvipsnames]{xcolor}
\usepackage{dcolumn} 
\newcommand{\mycomment}[1]{}

\definecolor{greybox}{RGB}{255, 99, 71}
\definecolor{redcircle}{RGB}{190, 10, 30}
\definecolor{apoor}{RGB}{255, 99, 71}
\definecolor{arich}{RGB}{65, 105, 255}

\definecolor{bimodal}{RGB}{255, 20, 147}
\definecolor{unimodal}{RGB}{138, 43, 226}
\definecolor{multimodal}{RGB}{30, 144, 255}
\definecolor{smeared}{RGB}{50, 205, 50}
\definecolor{GSE}{RGB}{255, 99, 71}

\definecolor{GSE1}{RGB}{100, 149, 237}
\definecolor{GSE2}{RGB}{60, 179, 113}
\definecolor{GSE3}{RGB}{255, 99, 71}

\usepackage{tikz}
\usetikzlibrary{shapes.geometric, shapes.symbols}

\newcommand{\hexamark}[1]{%
\tikz[baseline=(H.base)]
\node[draw, regular polygon, regular polygon sides=6, minimum size=3.5mm, inner sep=0pt, overlay] (H) {\tiny{#1}};%
}

\newcommand{\pentmark}[1]{%
\tikz[baseline=(P.base)]
\node[draw, regular polygon, regular polygon sides=5, minimum size=3.5mm, inner sep=0pt, overlay] (P) {\tiny{#1}};%
}

\newcommand{\starmark}[1]{%
\tikz[baseline=(S.base)]
\node[draw, star, star points=6, star point ratio=2.15, minimum size=4mm, inner sep=0pt, overlay] (S) {\tiny{#1}};%
}

\title[Timing the Gaia-Sausage merger]{The Last Galactic Firework: Timing the last significant merger with stars, globular clusters and $\omega$Centauri}

\author[C. F. P. Laporte \& M. D. A. Orkney ]{ Chervin F. P. Laporte$^{1,2,3,4}$\thanks{E-mail: chervin.laporte@obpsm.fr} and Matthew D. A. Orkney,$^{2,3}$ \vspace{0.1cm}\\
$^{1}$LIRA, Observatoire de Paris, Universit\'e PSL, Sorbonne Universit\'e, Universit\'e Paris Cit\'e, CY Cergy Paris Universit\'e, CNRS, 92190 Meudon, France\\
$^{2}$Institut de Ci\`{e}ncies del Cosmos (ICCUB), Universitat de Barcelona, Mart\'{i} i Franqu\`{e}s 1, E-08028 Barcelona, Spain\\
$^{3}$Institut d'Estudis Espacials de Catalunya (IEEC), E-08034 Barcelona, Spain\\
$^{4}$Kavli IPMU (WPI), UTIAS, The University of Tokyo, Kashiwa, Chiba 277-8583, Japan\\
}

\date{Accepted XXX. Received YYY; in original form ZZZ}

\pubyear{2025}

\begin{document}
\label{firstpage}
\pagerange{\pageref{firstpage}--\pageref{lastpage}}
\maketitle

\begin{abstract}
We present a robust method to empirically infer the timing of the last significant merger in the Milky Way which is tested against fully cosmological models of galaxy formation. We apply it to Milky Way subgiant stars with spectro-photometric ages, finding that the last significant merger (Gaia-Sausage-Enceladus, GSE), occurred $\sim11\,$Gyrs ago. This coincides with the birth of a coeval in-situ group of globular clusters (GCs), which constrains the merger-induced starburst (hereafter {\it Tain\'{a}}) to have occurred at $11.2\pm 0.1\,\rm{Gyr}$, the most precise dating of this merger event.
The GSE's most metal-rich GCs were also born around this time ($\tau=10.9\pm0.1\,\rm{Gyr}$) and likely formed during the merger interaction prior to disruption of the GSE. We argue that $\omega$ Centauri is the most likely candidate for the surviving remnant of the GSE, and show that its stellar populations have final ages and metallicities consistent with the GSE GCs together with observational evidence it may have been affected by bar resonances.
Furthermore, we argue that the mean metallicity for which stellar orbits transition from halo-like to disc-like kinematics shows an upward inflexion point at $[\rm{Fe/H}]\sim-1.33$, and this sets an upper-limit for the age when the disc was forming. To corroborate this, we identify proto-MW GCs with highly disc-like orbits that formed before the last significant merger (with ages up to $\tau=13.0\pm0.5\,\rm{Gyr}$). This places the disc formation time as far back as as $z_{\rm disc\, form}\gtrsim4$.


\end{abstract}




\section{Introduction} \label{sec:introduction}

In the current $\Lambda$CDM cosmological model \citep{Planck2014}, galaxies grow hierarchically \citep{efstathiou1985} through the condensation of gas inside dark matter haloes \citep{white1978}. Traces of this hierarchical growth can often be seen in their stellar halos, which form primarily from disrupted accreted systems \citep{bullock05}. At the scale of the Milky Way (MW), the $\Lambda$CDM framework predicts that the vast majority of accreted stars in the stellar halo were formed within a relatively few massive satellites \citep{robertson05, monachesi2019}, typically 1~to~3 within $20\,\rm{kpc}$ of the Galactic centre \citep{fattahi20}. A direct corollary from this should follow: the age--metallicity relation of the stellar halo should be dominated by its most massive contributor. This should set the age limit for the last significant merger. 

Over time, a disc galaxy will gradually build up its angular moment from high-redshift to the present-day. This can be witnessed in its present-day distribution of orbital circularity (or rotational velocity) monotonically rising as a function of metallicity \citep{belokurov2022}. In the MW, this so-called spin-up marks a region where the transition between velocity dispersion support and rotational support rises significantly. 
The sharp transition in the MW may not signal a rapid formation process \citep[as argued in][]{belokurov2022}, but rather reflect how the last significant merger can shape this curve \citep{chandra2024, orkney26}. Thus, one may draw a direct link between the time of pericentric passage and the spin-up metallicity. This in turn, through the age--metallicity relation of the stellar halo can be used to deduce the time of such an impact.

With the advent of Gaia's first data release and following releases, it has become clear that the most massive contributor to the inner-stellar halo is the debris of the Gaia-Sausage-Enceladus\footnote{Earlier signs of the GSE have been reported in the early 2000s \citep[e.g.][]{chiba2000,freeman2002,brook2004,meza2005} as well as in \cite{deason13} through the existence of a break in the stellar halo profile of the MW.} (GSE) galaxy which makes up to more than 50\% of the mass budget within 20 kpc \citep[e.g.][]{belokurov2018, haywood2018,helmi2018, naidu2020}. The next contributor is Sgr with only $2\times10^{8}\,\rm{M_{\odot}}$ in stellar mass \citep{niederste-ostholt10}. In this work, we are interested in constraining the time of the last significant merger (the GSE). To this end, we make use of two datasets: MW field subgiants from \citep{xiang2022} and Galactic globular clusters (GCs) from \citep{vandenberg2013}. The subgiant ages provide a powerful measure of the age--metallicity relation in the stellar halo and the kinematics of the disc and stellar halo. Meanwhile, the GC ages provide an independent method for dating the merger, and with a promise of even greater precision. 

In section~\ref{sec:methods}, we describe the datasets used in this study and methods for deriving other quantities (e.g. circularities, energies). In section~\ref{sec:results} we present our results. We discuss our findings in the context of the formation of the MW in section~\ref{sec:discussion} and conclude in section~\ref{sec:conclusion}.


\section{Methods} \label{sec:methods}

In this study, we use two catalogues. For globular clusters (GCs), we make use of the compilation from \cite{vandenberg2013} which includes ages and metallicities for 55 Galactic GCs. We further cross-match this catalogue with that of \citet{vasiliev21} to obtain positions, line-of-sight velocities and proper motions for the GCs from Gaia eDR3. For stars, we use the \cite{xiang2022} catalogue of LAMOST subgiants with Gaia DR3 proper motions and photo-spectroscopically derived stellar ages. We keep stars with fractional age uncertainties of $\tau/\sigma_{\tau} \leq 10$. We assume that the distance from the Sun to the Galactic center is $d_{\odot}=8.122\,\rm{kpc}$ \citep{gravity18}, the solar height about the midplane is $z_{\odot}=20.8 \, \rm{pc}$ \citep{bennet19}, and that the solar motion relative to the Galactic center is $(U, \, V,\, W)_{\odot}=(12.9, \,245.6,\, 7.78) \,\rm{km\,s^{-1}}$ 
\cite{schoenrich10, reid14}. Energies and angular momenta are calculated within the MW potential model of \cite{bovy2015}, modified following the prescription described in \citep{belokurov2023}. We additionally calculate orbital circularities $\eta=L_{z}/L_{\rm circ}$, where $L_{z}$ is the vertical component of the angular momentum vector of a star/GC and $L_{\rm circ}$ the corresponding angular momentum for a stars/GC of energy $E$ on a circular orbit.

\section{Results} \label{sec:results}

\begin{table}
\setlength{\tabcolsep}{8pt} 
\begin{tabularx}{\columnwidth}{cXXXX}
\toprule
Symbol & Name & Age/Gyr & $\rm{[Fe/H]}$ & log(M/$\rm{M_{\odot}}$) \\
\midrule

\multicolumn{5}{l}{{\it Tain\'{a}} GCs} \\
\starmark{0} & NGC 5927 & 10.75$\pm$0.38 & -0.29& 5.63 \\
\starmark{1} & NGC 6304 & 11.25$\pm$0.38 & -0.37& 5.15 \\
\starmark{2} & NGC 6352 & 10.75$\pm$0.38 & -0.62&4.82 \\
\starmark{3} & NGC 6366 & 11.00$\pm$0.50  & -0.59& 4.53\\
\starmark{4} & NGC 6496 & 10.75$\pm$0.38 & -0.46& 5.11\\
\starmark{5} & NGC 6624 & 11.25$\pm$0.50 & -0.42& 5.23\\
\starmark{6} & NGC 6637 & 11.00$\pm$0.38  & -0.59& 5.29 \\
\starmark{7} & NGC 6652 & 11.25$\pm$0.25 & -0.76& 4.90\\
\starmark{8} & NGC 6838 & 11.00$\pm$0.38  & -0.82& 4.48 \\[2mm]
\multicolumn{5}{l}{Proto-MW GCs} \\
\hexamark{9}  & NGC 104  &11.75$\pm 0.25$& -0.76& 6.0 \\
\hexamark{10} & NGC 6171 &12.00$\pm 0.75$ & -1.03& 5.08 \\
\hexamark{11} & NGC 6218 &13.00$\pm 0.50$ & -1.33& 5.16 \\
\hexamark{12} & NGC 6362 &12.5$\pm 0.25$ & -1.07& 5.01 \\
\hexamark{13} & NGC 6717 &12.5$\pm 0.50$ & -1.26& 4.50 \\
\hexamark{14} & NGC 6723 & 12.5$\pm 0.25$& -1.10 & 5.37 \\[2mm]
\multicolumn{5}{l}{GSE GCs} \\
\pentmark{0} & NGC 362  & 10.75$\pm$0.25  & -1.30& 5.61 \\
\pentmark{1} & NGC 1261 & 10.75$\pm$0.25  & -1.27& 5.35 \\
\pentmark{2} & NGC 1851 & 11.00$\pm$0.25   & -1.18& 5.57 \\
\pentmark{3} & NGC 2808 & 11.00$\pm$0.38   & -1.18& 5.99 \\
\bottomrule
\end{tabularx}
\caption{List of subset of GCs of particular interest: {\it Tain\'{a}} starbust associated MW GCs, proto-MW GCs and GSE most metal-rich GCs. We also list their metallicities and ages as reported in \citet{vandenberg2013} and their masses \citet{Kruijssen2019} which used the absolute magnitudes of \citet{harris96} and assumed $M_{V,\odot}=4.83$ and $M/L_{V}=2$}
\label{tab:mergers}
\end{table}

\begin{figure*}
\centering
  \setlength\tabcolsep{2pt}%
    \includegraphics[keepaspectratio, trim={0.2cm 0.0cm 0.2cm 0.0cm}, width=\linewidth]{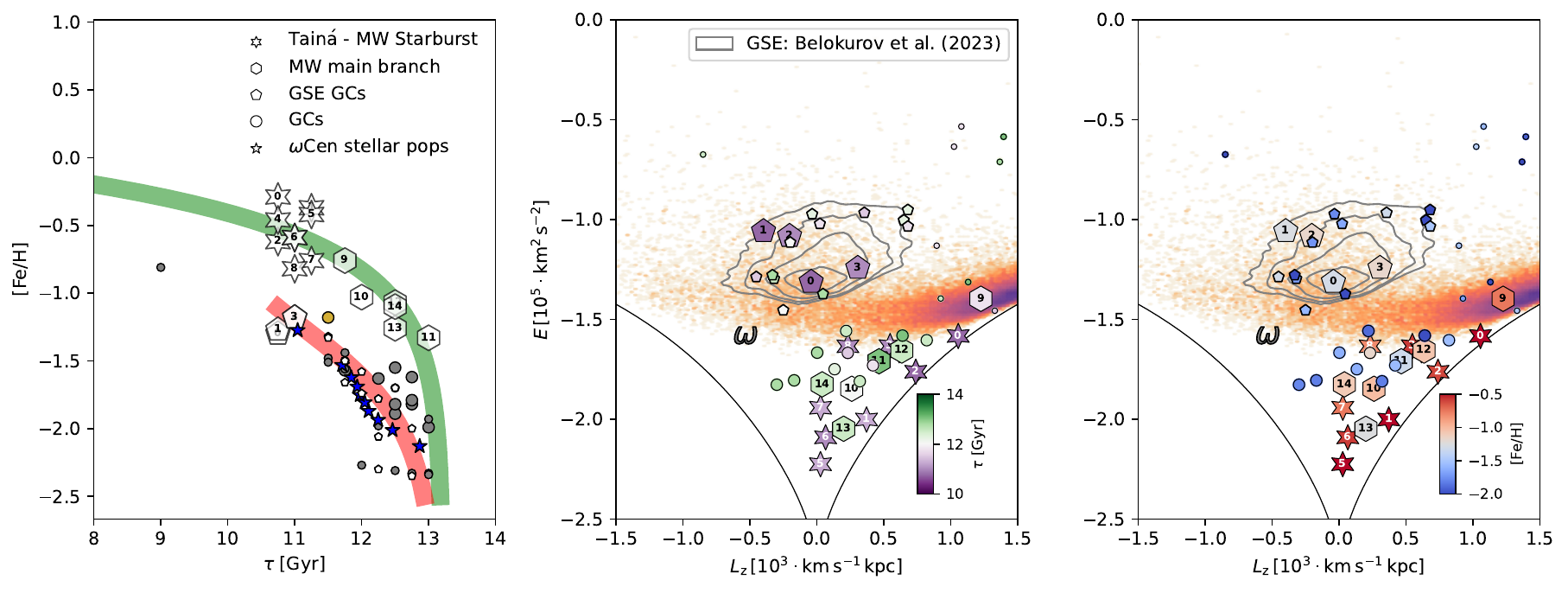}\\
\caption{
\textit{Left panel}: Age--metallicity relation for the \citet{vandenberg2013} GCs. Large hexagons and stars correspond to MW-branch GCs while pentagons highlight the most metal-rich GCs associated with the GSE. White pentagons represent other GSE-related candidate GCs identified in the vicinity of the GSE debris of \citep{belokurov2020}. Other inner-MW GCs with $E\leq-1.5\times10^{5}\,\rm{km^{2}\,s^{2}}$ are highlighted as large filled grey circles while outer GCs ($E>-1.5\times10^{5}\,\rm{km^{2}\,s^{2}}$) are represented by smaller filled ones.  The cluster sitting at $9\,\rm{Gyr}$ is Pal 12 which is associated to the Sagittarius dwarf galaxy \citep{ibata94}. Age--metallicity relations for the MW and GSE are shown as the thick green and red bands respectively, these are meant to guide the eye on the two sequences. NGC 6121 is highlighted in gold-yellow because its chemistry is consistent with it being in-situ. We will argue that it is most likely a MW-branch GC in section~\ref{sec:discussion}. Mean ages of stellar populations of different metallicity bins in the nuclear star cluster $\omega$Centauri ($\omega$Cen) from \citet{clontz24} are shown as blue stars. 
\textit{Middle panel}: Energy versus angular momentum plane. The background density map corresponds to the distribution of subgiants from \citet{xiang2022}. Despite being a survey in the Northern hemisphere, LAMOST has numerous stars with guiding radii satisfying $0<R_{\rm g}<5$, enabling it to probe the formation of the inner MW. MW-branch GCs colour-coded by their ages are mostly confined to the inner MW ($R_{\rm g}<5\,\rm{kpc}$), with the exception of NGC 104 which has a guiding radius of $R_{\rm g}\sim8\,\rm{kpc}$. The most metal-rich GSE GCs cluster are not surprisingly in the region where the GSE is most prominent in Gaia DR3 \citep[see][]{belokurov2020, orkney26}.
We also highlight the other GSE-related GCs as pentagons. $\omega$Cen's location is $E-L_{z}$ is also highlighted. 
\textit{Right panel}: Same as the middle panel, except that the GCs are colour-coded according to their metallicity.} 
\label{fig:age_feh}
\end{figure*}

In the left panel of Figure \ref{fig:age_feh}, we present the age--metallicity relation for the GCs in \cite{vandenberg2013}. The presence of two sequences (one belonging to the MW in-situ population, the other mostly to the ex-situ one\footnote{As we will see later, this sequence is not entirely ex-situ, particularly in the case of NGC 6121 (M4) and particularly in the central 5\,kpc region of the MW below metallicities of $[\rm{Fe/H}]$ where both sequences start to converge.}) has long been recognised \citep[e.g.][]{marin-franch09, forbes10, leaman13} which begins to separate clearly around [Fe/H]$\sim-1.5$ at the oldest ages. We separate GCs into different categories whenever possible, in particular unambiguous GCs belonging to the in-situ and GSE population which are represented by numbered symbols while the other clusters are represented by circles. As it will become clearer throughout our analysis, we separate the in-situ population into a starburst ({\it Tain\'{a}\footnote{From the Tupi-Guarani origin word signifying ``star'' or ``morning star'', marking this new rebirth phase in the lifetime of the Galaxy.}}) and proto-MW branch population, which are depicted by stars (numbered from 0 to 8) and hexagons respectively (numbered from 9 to 14). An additional 4 long known co-eval GCs belonging to the GSE are also highlighted as pentagons, numbered from 0 to 3. These are listed in Table \ref{tab:mergers} with some auxilary properties (mass, ages and metallicities). Other likely GSE-related GCs are also highlighted as smaller white pentagons. The remaining GCs are represented as grey circles which are split in two sizes depending on whether their energy is higher or lower or equal than $E=-1.5\times10^{5}\,\rm{km^{2}\,s^{-2}}$ respectively. This split is only to guide the eye and distinguish between tightly bound GCs (comprising of in-situ proto-MW GCs and perhaps some accreted ones) and loosely bound GCs (mostly comprising of accreted GCs).

 To guide the eye, we overlay two chemical evolutionary tracks for the host MW and stellar halo using parameters corresponding to those of the Galaxy and GSE stellar masses using the leaky-box chemical evolution model parametrisation of \cite{massari2019,massari26}. In this model, the GSE follows a constant star formation rate and the age--metallicity evolves according to

 \begin{equation}
 t(x)=t_{f} + \Delta t \, e^{\frac{-10^{x}}{p}},
 \end{equation}
where $t_{f}$ is the time where star formation shuts down, $\Delta t=t_{i}-tf$ the interval span of stars formation, with $t_{i}$ marking the beginning of star formation, $x=[\rm{Fe/H}]$ is the metallicity and $p$ is the effective chemical yield taken to be constant following the scaling relation of \cite{dekel03}
\begin{equation}
p=0.005 \left(\frac{M}{10^{6}\,\rm{M_{\odot}}}\right)^{0.4}.    
\end{equation} For the MW, we assume an exponentially declining star formation rate (i.e. $\rm{SFR}\propto e^{-\beta t}$) which proceeds over long timescales ($t_{i}\ll t_{f}$) where the age--metallicity relation follows the relation 
\begin{equation}
    t(x) = t_{i} - \frac{10^{x}}{\beta p}.
\end{equation}
For the MW we adopt a mass of $M_{*}=6\times10^{10}\,\rm{M_{\odot}}$ and $\beta=0.3$ and $t_{i}=13.2\,\rm{Gyr}$, whereas for the GSE we take $M_{*,\,\rm{GSE}}=4\times10^{8}\,\rm{M_{\odot}}$, $t_{i}=13.1\,\rm{Gyr}$ and $t_{f}=10.0\,\rm{Gyr}$.

Mean ages for stellar populations of different metallicities in the nuclear star cluster $\omega$Centauri ($\omega$Cen) derived by \cite{clontz24} are also overlaid as blue stars. Interestingly, we note that $\omega$Cen's  most metal-rich population not only shares the same age and metallicity as the GSE's 4 most metal-rich clusters (NGC 362, NGC 1261, NGC 1851, NGC 2808), but also that the GSE's GCs (as defined by association with its stellar debris contours in $E$ and $L_{z}$ \citep{belokurov2020}) closely follow $\omega$Cen's age--metallicity relation. This signals both a tight chemical and temporal link between the GSE's GCs and $\omega$Cen, reinforcing the idea the latter constitutes the remnant surviving core of the GSE \citep[e.g.][]{freeman2002,meza2005,limberg2022}.

\begin{figure}
  \setlength\tabcolsep{2pt}%
    \includegraphics[keepaspectratio, trim={0.0cm 0.5cm 0.0cm 0.0cm}, width=\columnwidth]{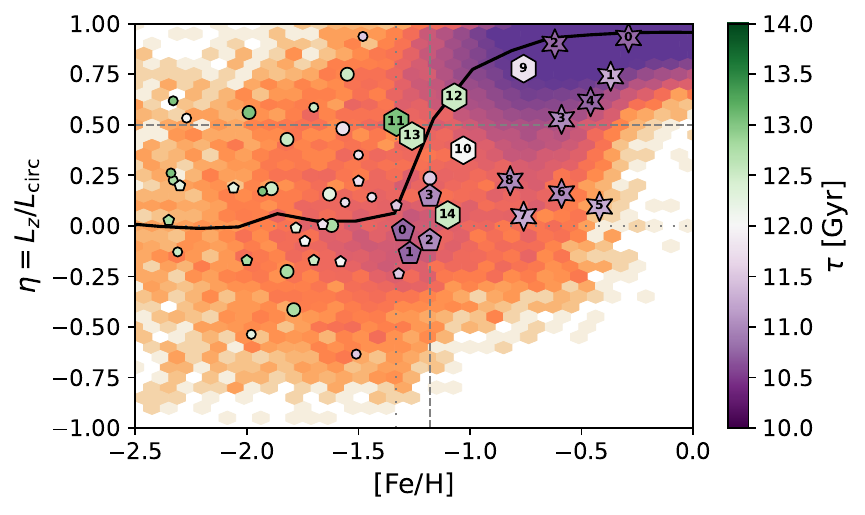}\\
\caption{Circularity versus metallicity diagram for stars and GCs in the MW. The solid black line marks the median curve for the stars. As expected, the curve flattens to $\eta\sim0$ at the metal-poor end ($-2.5<[\rm{Fe/H}]<-1.5$), consistent with a velocity dispersion dominated system. Around $[\rm{Fe/H}]\sim-1.4$ we notice a sharp inflection point where the curve abruptly rises and reaches $\eta=0.5$ at a metallicity of $[\rm{Fe/H}]_{\rm{spinup}}=-1.2$, which marks the spinup. This eventually plateaus about $\eta\sim1$ by $[\rm{Fe/H}]=-0.5$, at which point the system is fully rotationally supported. The 4 GSE most metal-rich GCs are as expected clustered around $\eta\sim0$. The starburst MW GCs span a wide range of circularities as expected from galaxy formation models \citep{orkney26}. The proto-MW GCs with ages $\tau>11\,\rm{Gyr}$ have circularities ranging $0<\eta<0.8$. Such a distribution clearly testifies that prior to the GSE merger the MW was already a rotationally supported system.}
\label{fig:circularity}
\end{figure}

Figure \ref{fig:circularity} shows the circularity versus metallicity plane for the MW with the GCs overplotted. The black line marks the median circularity relation from which the spin-up metallicity can be read-off ($[\rm{Fe/H}](\text{spin-up})=-1.2$). This value has typically been interpreted as the disc formation age, and the spread about this value to represent the duration of disc formation. However, this interpretation is not adequate, since {\it the present-day distribution is not synonymous with the distribution at birth}. Indeed, in the high-$z$ Universe, galaxies are susceptible to undergo mergers which will modulate their kinematics at birth \citep{McCluskey2024}. These can for example heat disc stars to the point that they end up with halo-like kinematics \citep{purcell2010} and have been recently identified in the MW \citep{dimatteo2019, belokurov2020} as the {\it Splash} population. 
Thus, a more accurate interpretation of the spin-up is the point where the MW underwent its last significant merger \citep[see][]{chandra2024, orkney26}.

In cosmological simulations, the spin-up in the metallicity versus circularity relation is initially less abrupt and shows significant rotation at lower metallicities when compared to the relation at present-day \citep{chandra2024, orkney2025}. The same is likely to hold for the MW. In this case, we note that the metallicity where the curve asymptotes to its $\eta=0$ value ($[\rm{Fe/H}]_{\rm{disc\,onset}}\sim-1.33$) should in fact set an upper-limit for the onset of disc formation. 

Interestingly, the oldest in-situ GC in the disc (NGC 6218) shares this same metallicity and formed at $\tau_{\rm discy}=13\pm 0.5$ Gyr. In other words, this is a clue that NGC 6218 formed within a proto-disc that was already spinning by $z=4$. Unfortunately, field stars cannot be used to anchor this age further due to the large error uncertainties which soar with increasing age. However, as stellar ages improve, so will our chemo-kinetic diagnostic of age-dating the onset of disc formation.

\begin{figure}
  \setlength\tabcolsep{2pt}    \includegraphics[keepaspectratio, trim={0.0cm 0.5cm 0.0cm 0.0cm}, width=\columnwidth]{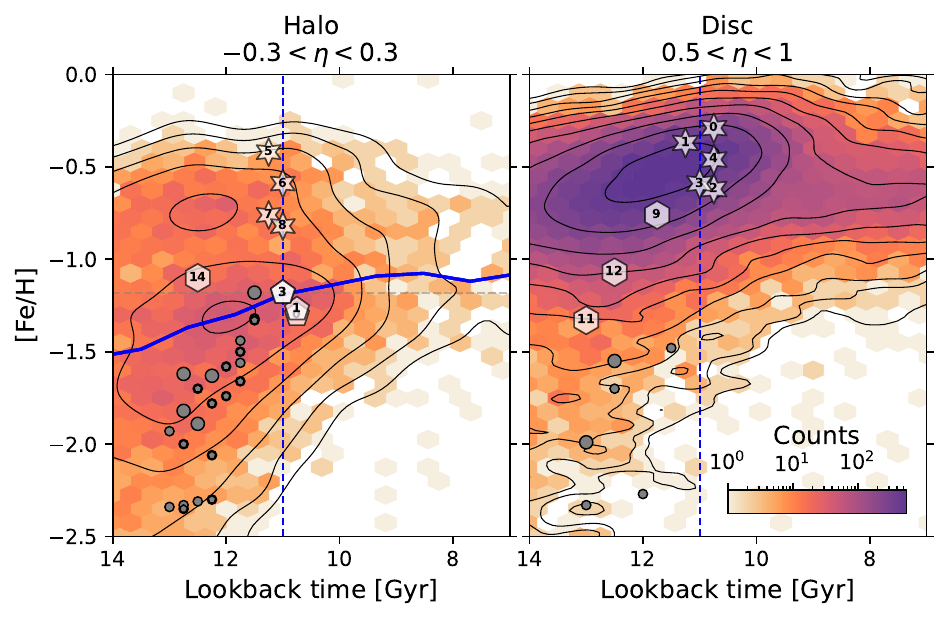}\\
\caption{Age--metallicity relations for the MW. {\it Left panel}: Age--metallicity relation for a sample of stars (background hexabinned data) and GCs (symbols similar to Fig. 1) with halo kinematics defined as $-0.3<\eta<0.3$. The horizontal dashed line marks the spinup metallicity $[\rm{Fe/H}]=-1.2$. The thick green line is the median of the relation and its intersection with the spinup metallicity marks the time of the pericentric passage of the last significant merger (i.e. the GSE) at $\tau_{\rm{peri}}\sim 11\,\rm{Gyr}$. This time coincides also with the age of the starburst GCs and 4 of the most metal-rich GCs in the GSE (including the age of the most metal-rich populations of $\rm{\omega}$Cen). {\it Right panel}: Age--metallicity relation of a sample of disc stars and GCs defined as $0.5<\eta<1$. Again, about half the starburst GCs have large circularities suggesting that these were already forming in a rotationally supported system. This is further corroborated by the fact that the pre-starburst GCs also have large circularities suggesting that the high-alpha disc was already in the process of formation at $z\geq 4$.}
\label{fig:age-metallicity-comp}
\end{figure}

When looking at both the age--metallicity relation from the GCs and that of the stars, as shown in Fig \ref{fig:age-metallicity-comp}, one notices that stars have shallower slopes than the sequence delineated by GCs. This is expected because GC ages are precise to $\sigma/\tau\sim0.05$, whereas the uncertainty in stellar ages scales as $\sigma/\tau\sim0.1$ \citep{xiang2022}, inducing an artificially flattening in their age--metallicity relation which becomes significant at high lookback time. GC sequences in age--metallicity diagrams should in essence provide a more accurate and precise depiction of the chemical evolution in the host than field stars would, with the only downside being that GCs are a porous tracer of Galactic chemical evolution. In the left panel of Figure \ref{fig:age-metallicity-comp}, we calculate the median relation for the age--metallicity relation of field stars with halo kinematics ($-0.3<\eta<0.3$) to derive the age of the spin-up metallicity $[\rm{Fe/H}]=-1.2$. This gives an age of $\tau_{\rm spinup}\simeq11\,\rm{Gyr}$. We make use of the stellar halo here because in our interpretation, the spin-up metallicity is not causally connected to the formation and evolution of the disc but rather to the last significant merger with a galaxy which in the high-$z$ Universe was starforming and ended up being quenched shortly after interaction. As the name implies, the last significant merger should set the age--metallicity of the stellar halo (last) and dominate in mass (significant) \citep{robertson05}. In Appendix \ref{AppendixA}, we present tests of this method on data from cosmological hydrodynamical simulations of MW-like galaxies, which take into account the hierarchical assembly and build-up of stellar halos in $\Lambda$CDM.

Next, we note that a subset of the MW GCs (both with halo and disc-like orbits) line up remarkably well with our timing of the merger as derived by the spin-up metallicity. In \citet{orkney26} we proposed that these GCs formed from the ambient ISM of the proto-MW, triggered by the impact of the GSE (a merger-induced starburst). The GCs coincident with the time of merger and thus starburst are the following: NGC 5927, NGC 6304, NGC 6352, NGC 6366, NGC 6496, NGC 6624, NGC 6637, NGC 6652, and NGC 6838. We observe a broad range of orbital circularities for these GCs (see Fig \ref{fig:circularity}) in the range ($0.0<\eta<1.0$). This distribution is expected in gas-rich mergers akin to the GSE from cosmological simulations which predict a strong central starburst phase in the inner region of the host galaxy \citep{grand2018,orkney2022, orkney2025} involving star-forming gas with a wide range of kinematics with $-0.3<\eta<1.0$. This is due to the fact that during a violent merger collision, the host's star forming gas loses angular momentum to feed the central parts of the galaxy leading to GCs forming on both halo-like and disc-like orbits. As already noticed in \cite{orkney26}, these GCs all share the same metallicity spread as the ISM, such that they outline that of the the chemical high-$\alpha$ disc. This spread of $0.5\,\rm{dex}$ also agrees with expected values from idealised experiments of star formation in turbulent high-$z$ gas-rich discs \citep[e.g. see Fig. 14 of][]{bland-hawthorn25}.


 We can use these ages to collectively derive an age for the starburst. Taking the weighted average of these GC ages, we derive a time for the onset of the starburst of $\tau_{\rm merger} \simeq \tau_{\text{Tain\'{a}}}=11.2\pm0.1\,\rm{Gyr}$. This places the merger at a redshift of $z_{\rm{GSE \,merger}}\sim2.5$. To our knowledge, this is the strongest constraint on timing the GSE-merger. The strength of the method lies in the corroboration of results through two independent means, using both field stars through photo-spectroscopically derived ages as well as the ages of individual clusters. Another remarkable feature is the co-evalness of the youngest GCs associated with the GSE \citep[see][]{massari2019, boldrini25} with those of the MW starburst population. This is consistent because star formation in the GSE would have been rapidly quenched after the collision \citep[e.g.][]{amarante2020}.
 
 These GSE GCs which also happen to be the most clearly identifiable ones due to their high degree of clustering in the space of energy angular momentum (see Fig \ref{fig:age_feh}), as well as circularity versus metallicity planes (Fig \ref{fig:circularity}): NGC 362, NGC 1261, NGC 1851, NGC 2808 (annotated following the same order). They also happen to be the most metal-rich GSE GCs with metallicities matching that of the spin-up (see Fig \ref{fig:circularity}).
 \footnote{We note that there are other more metal-rich ex-situ GCs but these belong to the Sagittarius dwarf galaxy which is known to have been accreted onto the MW at later times.} The weighted mean age of these four GSE GCs $\tau_{\rm GSE \, starburst}=10.9\pm 0.1\,\rm{Gyr}$ coincide very well with the time of the merger derived earlier.

The idea of a host-and-satellite mutual starburst has not been discussed much in the context of the MW and GSE; however, this is not a new idea and was in fact suggested as a mechanism behind the formation of multiple stellar populations in some dwarf galaxies through gas compression via ram-pressure around pericentric passage \citep{genina19}. In Appendix \ref{AppendixB} we use the {\sc auriga} simulations suite \citep{grand2024} to show that, for realisations involving MW-like and gas-rich GSE-like mergers, the merging satellite and host galaxy both undergo starbursts during the interaction. This further enforces the idea this scenario could have happened in our own MW, particularly with respect to the GSE merger.

 Finally, although the nuclear star cluster $\omega$Cen is not part of the \cite{vandenberg2013} catalogue, it is thought to be the remnant core of an ancient dwarf galaxy which merged with the Galaxy \citep[e.g.][]{freeman2002} possibly associated with the last significant merger \citep[e.g.][]{limberg2022}. We note that the age of its most metal-rich population, which has a median metallicity of $\rm{[Fe/H]}=-1.3$\footnote{While $\omega$Cen's MDF goes all the way to metallicities as high as $\rm{[Fe/H]}\sim 0.75-0.5$, this is an extreme tail of the distribution and \cite{clontz24} still show the median metallicity of this long tail to be skewed to $\rm{[Fe/H]\sim-1.3}$.} \citep[$\tau=11\,\rm{Gyr}$,][]{clontz24}, coincides very well with that of the most metal-rich GSE GCs.


Our method seems to tell a consistent story, but does it hold? To test this, we apply the same method to galaxies formed in cosmological simulations from the {\sc auriga} suite \citep{Auriga}, where we also convolved some age errors on the star particles proportional to those expected in the LAMOST catalog of \cite{xiang2022}. This is further discussed in Appendix \ref{AppendixA}, where we show that our method correctly recovers and infers the timing of the pericentric passage and subsequent starburst in the host galaxies, giving credence to our presented empirical method to infer the timing of the last significant merger.

\section{discussion} \label{sec:discussion}

Our estimates only considered GC data from \cite{vandenberg2013}. We deliberately did not use or mix data from other studies \citep[e.g.][]{forbes10, dotter11, Kruijssen2019} for a number of reasons which we list below. The data in \cite{forbes10} are a compilation of GC ages derived by different studies with different methodologies. As such, it does not represent a harmonised dataset. The ages in \cite{Kruijssen2019} are the averaged values from several different studies. However, we argue that this naturally introduces biases because the source measurements were not derived using similar methodologies. This leaves us with the ages derived by \cite{vandenberg2013}, which to this day remains the largest dataset with ages derived in a systematic consistent manner containing 55 clusters. Furthermore, many of the GCs in \cite{dotter10, dotter11} are also contained in the \citep{vandenberg2013} catalog.

Inspecting the \cite{pace25} local volume database, we noticed that most GCs with metallicities in the range $-0.8<\rm{[Fe/H]}<0.0$ are already included in the \cite{vandenberg2013} catalog, with a few exceptions (E3, IC 4499, NGC 6342, NGC 6388, NGC 6441, NGC 6528, NGC 6553). Interestingly, 5 of these belong to the inner-region of the MW (NGC 6342, NGC 6388, NGC 6441, NGC 6528 and NGC 6553) and 3 share similar ages to the ``Tain\'{a}'' GCs. These are NGC 6441 \citep[$\tau=11.26\pm0.90\,\rm{Gyr}$,][]{dotter11}, NGC 6528 and NGC 6553 \citep[both with $\tau=11.0\pm0.5\,\rm{Gyr}$,][]{ortolani25}. If we were to consider these as additional in-situ GCs, this would imply that the majority of the MW's surviving in-situ clusters were formed at a redshift of $z=2.5$. The tight age--metallicity sequence seen in these GCs is not a mere artifact; it is naturally expected for GCs that are born from a starburst. Moreover, the wide range of orbital circularities spanned by the GCs ($0\le \eta \le 1$) is consistent with predictions from numerical galaxy formation models of merger induced star formation \citep[][]{orkney26}.

\subsection{Kraken/Heracles?}
Our method has solely focused on the last significant merger, which we have associated with the GSE (which is the only ancient accretion event reported which has an unambiguous GC-stellar debris association \footnote{Other associations have been made however these are more debated such as the \textit{Sequoia} and Helmi streams \citep{helmi99, myeong19}.}), however, we showed in \cite{orkney2022} that it should be possible to disentangle even earlier merger through starburst signatures (which could be accompanied by the formation of GCs) in order to constrain the existence of the putative Kraken
/Heracles merger \citep{Kruijssen2019, massari2019, horta2021}. 
We do not attempt this here, as the LAMOST subgiant sample of stars does not probe well enough the inner-most regions of the MW. This may be probed in the future with surveys like 4MOST \citep{4MOSTs3} 

Recently, \cite{massari26} reported the existence of three separate GC sequences belonging to the in-situ Galaxy, the GSE and possibly the Kraken merger. However, their result rests entirely on postulating that NGC 6121 (M4) belongs to the Kraken galaxy, i.e. making it ex-situ. 
This scenario at face value, may seem plausible when considering solely the age and metallicity of this GC, but less so when accounting for its orbit (or internal chemistry). NGC 6121 is the most metal-rich GC that \citet{massari26} associate with Kraken, meaning it formed in the chemically enriched centre of its progenitor's potential well. As Kraken accreted onto the ancient MW, its orbit decayed to lower energies due to dynamical friction, accompanied by the tidal stripping of stars and GCs. The tightly-bound central regions of Kraken would have been among the last ones to get stripped. NGC 6121 should therefore reside at the lowest orbital energies within the Kraken debris footprint. However, NGC 6121 occupies the highest energies among the Kraken GC candidates, so this scenario is not physically viable from a dynamical standpoint.

\begin{figure}
  \setlength\tabcolsep{2pt}    \includegraphics[keepaspectratio, trim={0.0cm 0.5cm 0.0cm 0.0cm}, width=\columnwidth]{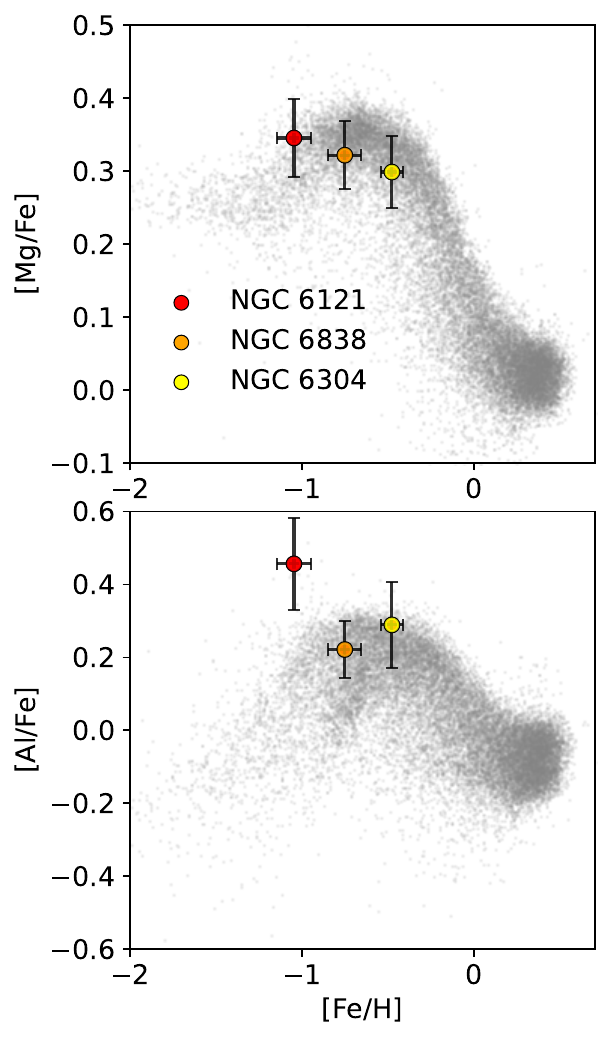}\\
\caption{{\it Top panel}: $[\rm{Mg/Fe}]$ versus $[\rm{Fe/H}]$ abundance diagram for APOGEE stars with energies $E<-1.5\cdot 10^{5}\,\rm{km^{2}\,s^{-2}}$ and individual cluster abundances for NGC 6121, NGC 6838 and NGC 304 in red, orange and yellow respectively. {\it Bottom panel}: $[\rm{Al/Fe}]$ versus $[\rm{Fe/H}]$ abundance diagram for APOGEE stars with energies $E<-1.5\cdot 10^{5}\,\rm{km^{2}\,s^{-2}}$ and individual cluster abundances for NGC 6121, NGC 6838 and NGC 304. All three clusters have abundances consistent with them being being in-situ according to the diagnostics from \citet{hawkins2015, belokurov2022}}
\label{fig:abundances_gcs}
\end{figure}

Aside from NGC 6121's derived age, there is also a chemical argument to be made about its nature. \cite{carretta13} report that NGC 6121 has a metallicity of $[\rm{Fe/H}]=-1.17$, a magnesium abundance of $[\rm{Mg/Fe}]= 0.55 \pm 0.01 $ and aluminium abundance of $[\rm{Al/Fe}]=0.58 \pm 0.03$. We can also compare these abundances with those in APOGEE together with the the added value GCs data from \citep{schiavon24}. In Figure \ref{fig:abundances_gcs}, we present $[\rm{Mg/Fe}]$ versus $\rm{[Fe/H]}$ and $[\rm{Al/Fe}]$ versus $\rm{[Fe/H]}$ abundances for stars with $E<-1.5 \cdot 10^{5}\,\rm{km^{2}\,s^{-2}}$ together with mean abundances for three clusters (NGC 6121, NGC 6838, NGC 6304). Chemically speaking, all three clusters have abundances consistent with them belonging to the high-$\alpha$ in-situ component of the Galaxy \citep[e.g.][]{hawkins2015,belokurov2022}.

\subsection{$\omega$Cen and its link to GSE through 1:1 bar resonance}

Finally, we note that $\omega$Cen's most metal-rich stellar population \citep[][]{clontz24} shares the same age as the four most metal-rich GCs of the GSE (see Figure \ref{fig:age_feh}). This signals that $\omega$Cen may very well be the nuclear star cluster of the GSE, as it seems highly unlikely that two separate galaxies with similar chemical evolutionary tracks could merge with the MW at the exact same time. This idea is not new \cite[see][]{freeman2002}. While a MW-like galaxy under the LCDM paradigm is predicted to be formed from 1 to 3 major building blocks \citep{robertson05, fattahi20}, it would be difficult for ancient GCs (i.e. those that formed before the accretion of the GSE) to have retained their skewed orbital circularity distribution ($\eta>0$) if they had been bombarded by 3 relatively significant mergers in the first 3 Gyrs of the MW's history. We thus argue that $\omega$Cen is most likely the nuclear star cluster of the GSE \citep{limberg2022}.



A common point of contention regarding association of $\omega$Cen to the GSE is that it is on a retrograde orbit in the MW, whereas most of the GSE debris is close to no net rotation. This has led some studies to suggest it may be related to another substructure, the {\it Sequoia} \citep{myeong19}. Whether the {\it Sequoia} is linked to the GSE event, or is its own independent substructure, is also not well established. A galaxy the mass of the GSE in the range $3\cdot 10^{8}<M/\rm{M_{\odot}}<1\cdot 10^{9}$ \citep[see][for various mass estimates]{belokurov2018, helmi2018, lane2023}, would have started out with a chemical gradient and dispersed its debris in non-trivial ways in the stellar halo \citep{jean-baptiste17} occupying a wide portion $E-L_{z}$ space. In fact, \cite{koppelman20} and \cite{amarante2020} present idealised N-body and hydrodynamical N-body models which show that \textit{Sequoia} could indeed form the metal-poor tail of the GSE's outer structure. The same has been shown also in cosmological simulations \citep[e.g.][]{orkney2023}. \citet{pagnini2025b} recently identified a number of MW GCs to share chemical similarities with $\omega$Cen and proposed the existence of a common link between them through an ancient dwarf progenitor {\it Nephele}. The authors do not claim this galaxy to necessarily be of separate origin to the GSE. However, given the argument we presented in \ref{fig:abundances_gcs}, it becomes increasingly difficult to argue for a distinct origin of {\it Nephele} as two (or three if counting {\it Sequoia}) galaxies of similar masses to the GSE could have contributed to the stellar halo at the {\it same} time. 

We argue that the constraints on the mass of the stellar halo \cite{deason2019}, ages of the stellar populations in GCs \citep{vandenberg2013}, the most metal-rich GSE GCs, $\omega$Cen \citep{clontz24} and a LCDM prior \citep{robertson05}, strongly disfavour the ``unique'' nature of {\it Sequoia}, {\it Nephele} and other debris sharing very similar properties to the GSE \citep[see the discussions in ][]{amarante2020}. From a purely chemical perspective, \cite{souza26} makes the clearest chemical argument for this.

As such, Occam's razor postulates that the most significant merger was the GSE (see \ref{fig:abundances_gcs} and \ref{fig:age-metallicity-comp}), its dispersal across $E-L_{z}$ was messy \citep[][]{koppelman20, amarante2020} and $\omega$Cen must have been its nuclear star cluster \citep[see][and the age--metallicity constraint from Fig. \ref{fig:abundances_gcs}]{freeman2002,limberg2022}. In this scenario the retrograde nature of the {\it Sequoia} is just a reflection of stripping of outermost material prior to radialisation of the Sausage \citep{amorisco2017}.


This leaves us, however, with an apparent dynamical puzzle. How did $\omega$Cen lose enough energy to fall at lower depths than the majority of the GSE debris and why is it still on a retrograde orbit? It is not clear how dynamical friction (DF) alone, would work to this end, given DF in a MW-like potential does not circularise orbits but rather radialises \citep[e.g.][]{vandenbosch99,amorisco2017,vasiliev2022}, so at face value $\omega$Cen ought to be at lower energies but with a near-zero angular momentum \citep[][]{freeman2002}. Therefore, a mechanism beyond DF must be invoked. One class of dynamical mechanisms capable of simultaneously changing the energy and angular momentum of a halo star (or any collisionless tracers, i.e. GCs) are bar-driven resonances \cite{dillamore2022,dillamore2023}.  Using cosmological simulations, \cite{tomlinson26} shows that the 1:1 resonance can take a star on slightly prograde orbit at high energy down to a retrograde one in the span of $1\,\rm{Gyr}$ (see their Fig. 6) with changes in energy and angular momentum similar as those required to shift a tracer along the locus connecting the central GSE debris \citep{belokurov2023} to $\omega$Cen  by $\Delta E\sim 5\cdot 10^{4}\,\rm{km^{2}\,s^{-2}}$ and $\Delta L_{\rm{z}}\sim 500 \rm{km\cdot s^{-1}\,kpc}$. We propose that this effect may have acted on $\omega$Cen to bring it at its current location in $E-L_{z}$ space with respect to the rest of the GSE debris. Thus, it may not be inconceivable that such a mechanism could have operated on $\omega$Cen at some point during the remaining $11\,\rm{Gyr}$ of its existence after the bar formed and evolved in the last $10-8\,\rm{Gyr}$ \citep{sanders24,haywood24} during which the pattern speed may have changed.

\begin{figure}
  \setlength\tabcolsep{2pt}    \includegraphics[keepaspectratio, trim={0.0cm 0.5cm 0.0cm 0.0cm}, width=\columnwidth]{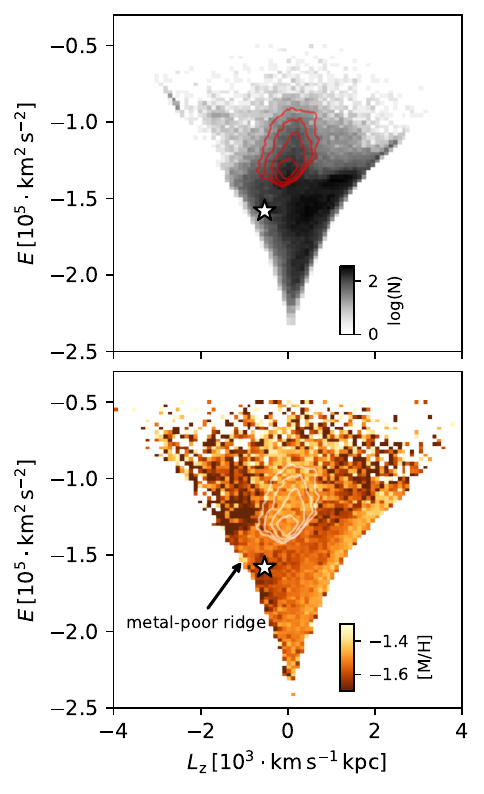}\\
\caption{$E-L_{\rm{z}}$ plane some with density contours for the GSE  where it is most prominently identified in the Gaia data \citep{belokurov2023} and location of $\omega$Cen (star). {\it Top panel}: Logarithmic star counts. {\it Bottom panel}: Same plane colour coded by metallicity $[\rm{ M/H}]$. The core of the GSE is more metal-rich than its surrounding by $\approx 0.25$\,dex. A diagonal metal-poor ridge connecting back to the GSE debris centroid bound by two more metal-rich regions can be seen with striking qualitative similarity with the prediction of bar driven resonance imprints on stellar halos in \citet{tomlinson26}. This may be tentative evidence for a bar driven resonance in the stellar halo which may have affected $\omega$Cen's past orbit.}
\label{fig:resonance_ridge}
\end{figure}

Particularly, \cite{tomlinson26} predict that such a resonance in the stellar halo in $E-L_{\rm{z}}$ would appear as a ridge which would be chemically more metal-rich than its surroundings. \cite{belokurov2022} show that on average field halo and in-situ component of the MW separate rather well at metallicities $-1.7<{\rm{[Fe/H]}}<-0.5$, however in order to detect substructure we need a sample larger than APOGEE. Fortunately, the \cite{li2024} catalog provides sufficient phase-space coverage in the Galaxy to attempt such a detection. In Figure \ref{fig:resonance_ridge}, we show the $E-L_{z}$ plane using the catalog from \citep{li2024}. For our selection, we make the following cuts: 

\begin{itemize}
    \item $\varpi/\sigma_{\varpi}>5$
    \item $\texttt{phot\_bp\_mean\_mag}$<16.5
    \item $\texttt{teff\_xp}<$5200
    \item $\texttt{logg\_xp}<$3.5
    \item $-1.9<\texttt{moh\_xp}<-1.3$
    \item $\texttt{aom\_xp}>0.1$
    \item $r/\rm{kpc}<1000$
\end{itemize}

The top panel shows the $E-L_{\rm{z}}$ plane in logarithmic star counts while the bottom panel shows it in the space of metallicity $[\rm{M/H}]$. We also overlay some density contours for the GSE  where it is most prominently identified in the Gaia data \citep{belokurov2023}. There it is evident that the core of the GSE is more metal-rich than its surrounding by $\approx0.25$\,dex. Furthermore, we note that this chemical map qualitatively resembles the prediction of hydrodynamical simulations of the GSE \citep[][see their Fig. 5]{amarante2020} which predict that the outer metal-poor regions of the GSE should spread about the extremities of the GSE central debris on both prograde/retrograde sides in $E-L_{\rm{z}}$ space. We also note a contribution of the in-situ component of the MW on the prograde side which is systematically higher in metallicity than the corresponding retrograde side of the $E-L_{\rm{z}}$ space. Unfortunately without $[\rm{Al/Fe}]$ abundances this slight in-situ contamination cannot be fully mitigated.

A diagonal metal-poor ridge connects back to the GSE debris centroid bound by two more metal-rich regions. Interestingly, the series of density contours of \cite{belokurov2023} to the right of the GSE centroid seem to also align parallel to the slope of the metal-poor ridge. This is qualitatively similar to the prediction of \cite{tomlinson26}. We suggest this may be tentative evidence for a bar driven resonance in the stellar halo which may have affected $\omega$Cen. A thorough investigation is beyond the scope of this contribution, but further exploration may be warranted with upcoming Gaia data releases and potentially 4MOST which we leave for future work.

\subsection{Timestamps in the formation of the early Milky Way}


In summary, both field stars and GCs paint a precise timeline for the formation of the proto-MW: high-$\alpha$ disc formation at $5<z<2.5$, the merger with the GSE at 11 Gyrs ago or $z\sim2.5$, and then the subsequent starbursts in the proto-MW and GSE progenitor galaxies. The merger also led to the formation of the {\it Splash} \citep{dimatteo2019, belokurov2020}, in which ancient proto-disc stars were liberated into the stellar halo or lost angular momentum. The abrupt drop in density of stars with metallicities in the range $-1<\rm{[Fe/H]}<-0.5$ with ages younger than $\tau=10\,\rm{Gyr}$ is a clear testament that the MW disc had a very quiescent later growth stage, only partially modulated by later accretion events such as the Sagittarius dwarf galaxy \citep{ibata94} and the Large Magellanic Cloud \citep[e.g.][]{weinberg06, laporte18}. 

Furthermore, our timing of the last significant merger ($\tau_{\rm merger} \simeq \tau_{\text{Tain\'{a}}}=11.2\pm0.1\,\rm{Gyr}$) consistently fits within the series of events which should have followed in the formation of the MW $\tau_{\rm disc \,formation}<\tau_{\rm merger}<\tau_{\rm \rm bar\, formation}$, whether one chooses $\tau_{\rm disc \,formation}\sim \tau_{\rm NGC\,6652}$ as we have argued \citep[see also][]{belokurov2024} or take the estimate from Mira variables placing disc formation around $z\sim3$ \citep{zhang24}. Indeed, bar formation is situated around $\tau_{\rm{bar \, formation}}\gtrsim 8 \,\rm{Gyr}$ \citep{sanders24,haywood24}.

\subsection{Outlook}
Although our technique is currently only applicable in the MW where we have exquisite 6D data, it could equally be extended to the extra-galactic realm in the not too distant future (e.g. with M31). Interestingly, it has recently been argued that Andromeda underwent a major merger some 2--3 Gyrs ago with an ancient galaxy \citep{hammer18, dsouza18}. Provided age estimates of GCs in M31 improve in the next years (e.g. with JWST), our method could presumably detect/infer even earlier mergers in Andromeda's history going as far back as cosmic noon.

\section{conclusions} \label{sec:conclusion}

In this contribution, we developed a novel method to probe the time of the last significant merger exploiting the tight link which should exist on the scales of MW-mass galaxies in $\Lambda$CDM between the circularity versus metallicity relation together with the age-metallicity relation of the stellar halo. We tested its efficacy on fully cosmological simulations of Milky Way-like galaxy formation and applied it on field stars and globular clusters (GCs) of the MW to show that: 

\begin{enumerate} 

    \item The last significant merger (GSE), happened $11.2\pm0.1\,\rm{Gyr}$ ago (see Fig. \ref{fig:age-metallicity-comp}). This is corroborated by two independent probes: field stars and a population of starburst globular clusters. \\
    
    \item This resulted in a starburst population of field stars and GCs ({\it Tain\'{a}}) : NGC 5927, NGC 6304, NGC 6352, NGC
6366, NGC 6496, NGC 6624, NGC 6637, NGC 6652, NGC
6838, NGC 6441, NGC 6528
and NGC 6553. \\

    \item During its last plunge towards the MW, the GSE underwent a simultaneous starburst at $\tau_{\rm GSE,\, starburst}=10.9\pm0.1\,\rm{Gyr}$ (see Fig. \ref{fig:age-metallicity-comp}).  These GCs (NGC 362, NGC 1261, NGC 1851, NGC 2808) share similar ages and metallicities ($[\rm Fe/H]\sim-1.3$) which also coincide with the most metal-rich stellar population of $\omega$Cen (see Fig. \ref{fig:age_feh}). We interpret this as a close chemical evolutionary link and association of $\omega$Cen with it being the nuclear star cluster of the GSE.\\
    
    \item The present-day orbital circularity distribution of field stars and GCs as a function of metallicity sets an upper-limit on the time of formation of the proto-disc and build-up of angular momentum prior to the GSE merger. This is situated at a time when the MW-progenitor reached a metallicity of $[\rm Fe/H]\sim-1.33$ which NGC 6652 anchors at $\tau=13.0\pm0.5\,\rm{Gyr}$ (see Fig. \ref{fig:circularity}), which at face value would place this around a redshift of $z_{\rm disc\,  form}\gtrsim4$.\\

    \item We argue that $\omega$Cen is the nuclear star cluster of the last significant merger progenitor galaxy (which we identify as the GSE). This is corroborated by the fact that the GSE GCs age--metallicity relation directly follows that of $\omega$Cen's stellar populations \citep[][see our Fig. \ref{fig:age_feh}, left panel]{clontz24}. Its location in the $E-L_{z}$ may indicate the effect of a possible 1:1 resonance with the bar which may have acted on it somewhere in the last $11 \, \rm{Gyr}$.\\

    \item We present tentative evidence of a metal-poor ridge of field stars in the stellar halo in $E-L_{z}$ connecting the GSE debris to lower energies retrograde orbits close to the vicinity of $\omega$Cen (Fig. \ref{fig:resonance_ridge}). This indicates that it may have been affected by a bar-resonance \citep[][]{tomlinson26}.
    
\end{enumerate}

\section{Acknowledgements}
 CL acknowledges useful discussions with Misha Haywood and Nils Hoyer. This work has used resources from the MareNostrum 4 supercomputer at the Barcelona Supercomputing Center (BSC). Further analysis and simulation work was performed on the NYX supercomputer at the Universitat de Barcelona (ICCUB). CL \& MO acknowledge funding from the European Research Council (ERC) under the European Union’s Horizon 2020 research and innovation programme (grant agreement No. 852839). CL also acknowledges funding from the Agence Nationale de la Recherche (ANR project ANR-24-CPJ1-0160-01). The author acknowledges financial support from the CEX2024-001451-M grant funded by MICIU/AEI/10.13039/501100011033. CL dedicates this to Thayenne.

\section*{Data Availability}

Any observational databases accessed in this work are publicly available through the provided citations. The {\sc auriga} simulation data is publicly available to download via the Globus platform as described in section~4 of \citet{Auriga}.

Other simulation data, and any post-processed data that is unique to this work, can be made available to the enquirer upon request.



\bibliographystyle{mnras}
\bibliography{references} 



\appendix

\section{Testing the method on cosmological simulations} \label{AppendixA}

In this section, we illustrate the success of our method on simulated data from the {\sc auriga} simulations suite. We refer to \cite{Auriga, grand2024} for a full description of the simulations, which comprise 30 MW-mass galaxies labeled Au-$i$, where $i = 1\dots30$. For brevity, we present results using just two realisations: Au-18 which has an accretion history closely resembling that of the MW with a GSE-like merger \citep{fattahi2019, orkney2025}, and Au-24 which has both a Kraken-like and GSE-like merger \citep{orkney2022}.

Our working assumption is that proto-discs in each galaxy spin-up early, and are then partially disrupted during the pericentric passage of the last significant merger. Therefore, the characteristic metallicity of stars at the onset of the spin-up (defined as $\eta(z=0)=0.5$) should correlate with the pericentre time via the stellar halo's age--metallicity relation.

\begin{figure*}
\centering
  \setlength\tabcolsep{2pt}%
    \includegraphics[keepaspectratio, trim={0.2cm 0.0cm 0.2cm 0.0cm}, width=\linewidth]{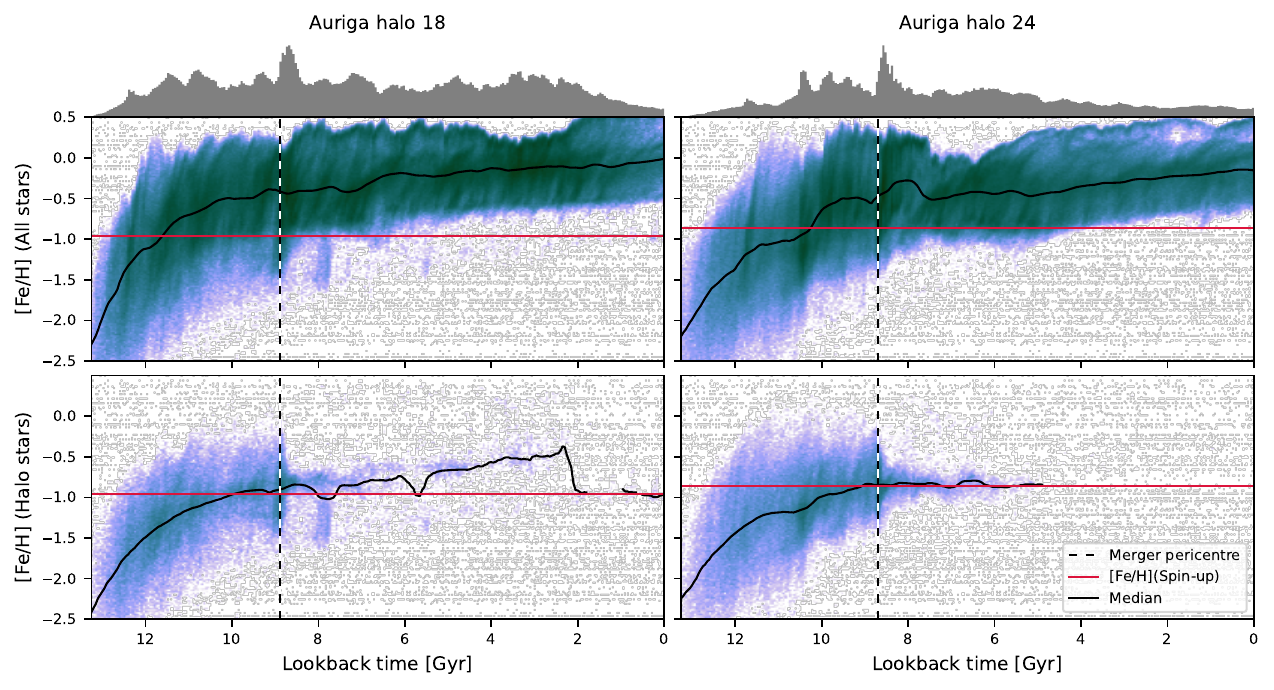}\\
\caption{
\textit{Top panels:} a 2D stellar mass-weighted histogram of the stars in each Auriga galaxy. The top axis includes the normalised 1D star formation rate in bins of 50\,Myrs. The black lines shows the running median, determined as the stellar mass-weighted median of $\pm5$ histogram columns (i.e. $\pm250\,$Myrs). The black-white dashed lines mark the pericentre time of the last significant merger in each simulation. The red lines mark the characteristic metallicity for which stars have a median orbital circularity of $\eta(z=0)=0.5$,  which we define to be the spin-up.
\textit{Bottom panels:} These panels show the same, but for a limited sample of stars on halo-like orbits $-0.3<\eta(z=0)<0.3$ and outside the disc plane $R_{z}>3\,\rm{kpc}$. This selection contains both in- and ex-situ stars, and the ex-situ portion hosts a high fraction of stars that came from the last significant merger ($49\%$ for Au-18 and $36\%$ for Au-24, where we would expect $\sim50\%$ for the MW).
}
\label{fig:test_tech}
\end{figure*}

Figure \ref{fig:test_tech} shows the age--metallicity relation for Au-18 (left) and Au-24 (right). The top panels show a stellar mass-weighted 2D histogram of the age--metallicity relation for all the stars within the galaxy at $z=0$. The 1D histogram on the top axis represents the overall star formation rate. The lower panels show the same but for a selection of halo stars with kinematics $-0.3<\eta(z=0)<0.3$, similarly as done with the observational data in section~\ref{sec:results}. The solid black lines show the median of the age--metallicity relation, which is what we use for our comparisons here.

Note that an intense in-situ starburst follows immediately after the pericentric passage of the last significant merger (white-black line). An intense starburst like this would provide an explanation for why the {\it Tain\'{a}} GCs sit on a mono-age line centred around the time of the inferred pericentric passage of the GSE (see Fig. \ref{fig:age-metallicity-comp}). Auriga lacks the fidelity to form star clusters of its own, so we cannot compare to the observational data in this case.

In both cases, the characteristic spin-up metallicity (red line) intersects the median stellar halo age--metallicity relation (black line) at a time close to the pericentric passage of the last significant merger.

We have further tested this method on data with a stellar age uncertainty of 10 per cent (comparable to observational limits) and still find that the method consistently retrieves the correct answer for the merger pericentre time. This gives us confidence that our method is well adapted for application in systems with MW-like accretion histories.

\section{Starburst in satellite and host} 
\label{AppendixB}

\begin{figure*}
\centering
  \setlength\tabcolsep{2pt}%
    \includegraphics[keepaspectratio, trim={0.2cm 0.0cm 0.2cm 0.0cm}, width=0.9\linewidth]{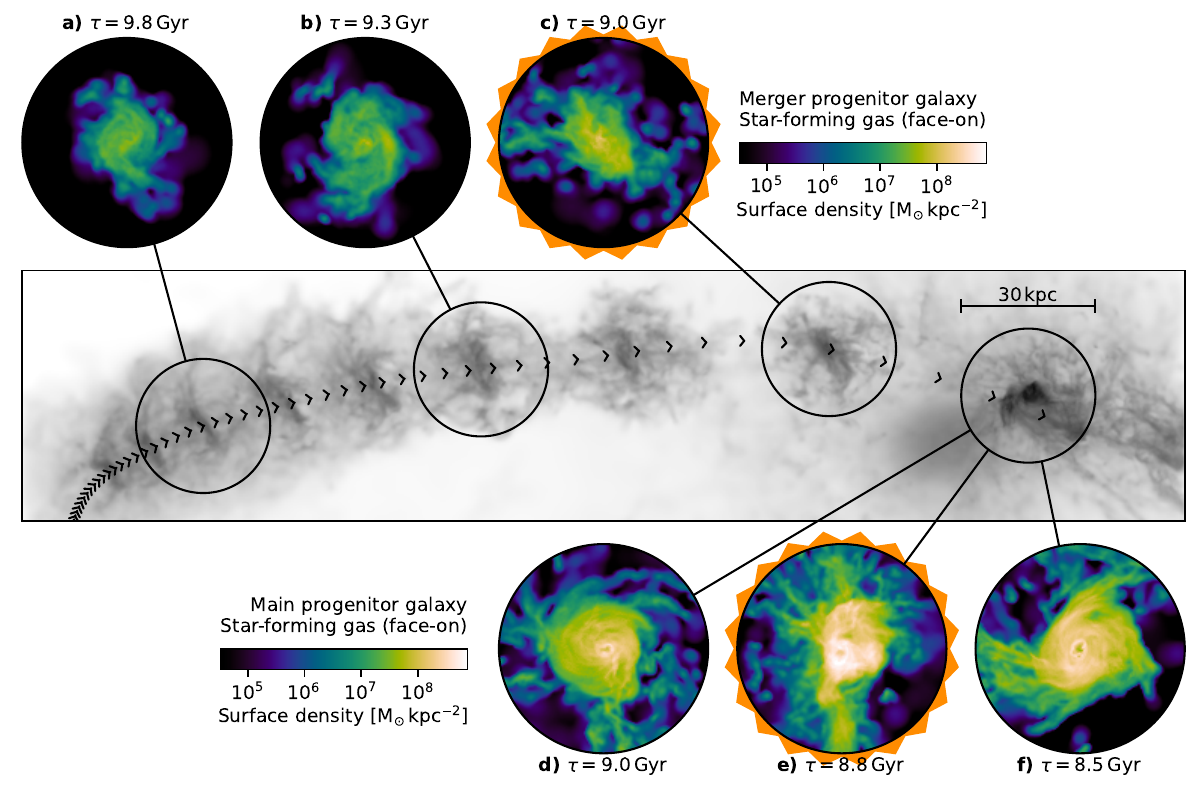}\\
\caption{An illustration of the last significant merger in Au-18. The black-and-white image shows the progression of the gas across a series of simulation snapshots, where the movement of the merger takes place from left-to-right. The simulation frame has been aligned on the orbital trajectory of the merger, such that any deviations above/below the $x$--$y$ plane are minimised. The final snapshot on the right-side of the image shows the colliding gas between the merger and host galaxies. The full merger path is indicated by the trail of black chevrons, for which the inter-chevron spacing gives an idea of the merger velocity. There are a series of circular zoom-in panels which show the star-forming gas within a 30\,kpc window in both the merger (top) and host (bottom) galaxies at a selection of lookback times. Instances where there is an elevated star formation rate are highlighted with an orange sunburst graphic. The peak gas densities achieved, in units of $\rm{M}_{\odot}\,\rm{kpc}^{-3}$, are as follows: \textbf{a)} $3.0\times10^7$, \textbf{b)} $6.3\times10^7$, \textbf{c)} $1.4\times10^8$, \textbf{d)} $8.0\times10^8$, \textbf{e)} $3.8\times10^9$, \textbf{f)} $1.8\times10^9$. This demonstrates how the merging galaxy undergoes a starburst during its final approach towards its orbital pericentre, whereas the host galaxy undergoes a much more intense starburst immediately after the merger pericentre.}
\label{fig:sf_sat_gal}
\end{figure*}

In this section we show that both the host and the merging galaxy can experience an elevated star formation rate as a result of their interaction. Figure \ref{fig:sf_sat_gal} illustrates this, where the black-and-white image shows the progression of the merger gas in a series of simulation snapshots transpiring from left-to-right.

As the satellite approaches pericentre (panel \textbf{c}, $\tau=9.0\,\rm{Gyr}$) we notice that the surface density of the central star forming gas increases by over half an order of magnitude. After the merger reaches pericentre at $\tau=8.9\,\rm{Gyr}$, the gas in the host galaxy experiences a considerable loss of angular momentum, which leads to a compaction of gas into the inner few kpc and a powerful starburst. A deeper analysis of the underlying physics is beyond the scope of this paper. However, previous studies have argued that ram-pressure compression can trigger star formation in the central regions of gas-rich satellites, while their outskirts are stripped \citep{genina19}. The existence of 4 metal-rich GSE GCs with similar ages to the MW starburst, suggests that such a mechanism could also have been at play during the last significant merger in our own Galaxy.

\label{AppendixC}




\bsp	
\label{lastpage}
1\end{document}